# Phase Change Observed in Ultrathin $Ba_{0.5}Sr_{0.5}TiO_3$ Films by *in-situ* Resonant Photoemission Spectroscopy

Y.-H. Lin

State Key Laboratory of New Ceramics and Fine Processing, Department of Materials Science and Engineering, Tsinghua University, Beijing, 100084, P. R. China

K. Terai[a]

Synchrotron Radiation Research Center, Japan Atomic Energy Research Insititute, SPring-8, Hyogo, 679-5198, Japan

H. Wadati, M. Kobayashi, M. Takizawa, J. I. Hwang, A. Fujimori[b]

Department of Physics and Department of Complexity Science and Engineering, University of Tokyo, Bunkyo-ku, Tokyo 113-0033, Japan

C.-W. Nan, J.-F. Li

State Key Laboratory of New Ceramics and Fine Processing, Department of Materials Science and Engineering, Tsinghua University, Beijing, 100084, P. R. China

S.-I. Fujimori, T. Okane, Y. Saitoh, K. Kobayashi

Synchrotron Radiation Research Center, Japan Atomic Energy Research Insititute, SPring-8, Hyogo, 679-5198, Japan

## ABSTRACT

Epitaxial $Ba_{0.5}Sr_{0.5}TiO_3$ thin films were prepared on Nb-doped $SrTiO_3$ (100) substrates by the pulsed laser deposition technique, and were studied by measuring the Ti $2p \rightarrow 3d$ resonant photoemission spectra in the valence-band region as a function of film thickness, both at room temperature and low temperature. Our results demonstrated an abrupt variation in the spectral structures between 2.8 nm (~7 monolayers) and 2.0 nm (~5 monolayers) $Ba_{0.5}Sr_{0.5}TiO_3$ films, suggesting that there exists a critical thickness for phase change in the range of 2.0 nm to 2.8 nm. This may be ascribed mainly to the intrinsic size effects.

**Key words**: $Ba_{0.5}Sr_{0.5}TiO_3$    thin film    resonant photoemission    phase change

Ferroelectric compounds with the perovskite $ABO_3$ structure, especially in thin film form, have attracted considerable attention due to their promising potential applications for alternative materials in ferroelectric random-access memories (FERAM) and metal-oxide semiconductor field-effect transistors (MOSFETs).[1] However, in previous experimental and theoretical investigations, a degradation of ferroelectric properties was observed with deceasing film thickness or particle size.[2-4] It is generally believed that there exists a critical size on the order of several to tens nanometers depending on materials, below which the spontaneous electrical polarization disappears. Up to now, understanding of what happens at this critical size is being intensively pursued through the combination of experimental and theoretical techniques. Fong and his co-workers[5] investigated $PbTiO_3$ thin films by a synchrotron x-ray scattering technique, and found that the ferroelectric phase existed in thickness down to 1.2 nm. Recently, Gerra *et al.*[6] reported that the critical thickness for ferroelectricity in ultrathin $BaTiO_3$ films with $SrRuO_3$ electrodes was 1.2 nm using a first-principles calculations. These results provided useful information for future microelectronic and micromechanical device applications with improved performance. However, it has been generally difficult to observe the critical size effect by conventional electric measurements due to large leakage current in the ultrathin films.[7]

In this work, we show that an *in-situ* resonant photoemission spectroscopy (RPES) monitoring technique is very useful to investigate the electronic structure of these ultrathin films and the critical size effect on the ferroelectric phase to paraelectric phase transition. It allows us to obtain high-quality thin films with desired thicknesses, and to maintain the chemical stoichiometric compositions and clean surfaces during the measurements. Additionally, because no electrodes are necessary for RPES measurements, one can study more intrinsic properties of ferroelectric thin films. Normally, bulk $(Ba,Sr)TiO_3$ changes from the paraelectric cubic (*m3m*) to ferroelectric tetragonal (*4mm*), orthorhombic (*mm*2), to rhombohedral (*3m*) phases as the temperature deceases.[8] Previous Ti $3p \rightarrow 3d$ RPES studies on Ti oxides have indicated that the strength of the resonant effect in each spectral feature varies across the valence band, correlated with the strength of the hybridized Ti 3*d* character into the predominantly oxygen 2*p* derived valence band.[9] Theoretical studies with first-principles calculations suggested that the hybridization between the Ti 3*d* and O 2*p* states should be essential for the ferroelectricity in BaTiO$_3$ and changes between the ferroelectric and paraelectric phases.[10] In particular, because one can enhance Ti 3*d* contributions out of the Ti 3*d*-O 2*p* hybridized valence band and the RPES spectra in the Ti 2*p* → 3*d* core-absorption region were reported to change across the ferroelectric phase transition,[11] the RPES spectra in the Ti 2*p* → 3*d* core-absorption region measured before and after the phase transition can be effectively used to investigate the critical size effect in (Ba, Sr)TiO$_3$ thin films.

Epitaxial Ba$_{0.5}$Sr$_{0.5}$TiO$_3$ (BSTO) thin films with various thicknesses (1.2 nm, 2.0 nm, 2.8 nm, and 200 nm) were deposited on Nb-doped SrTiO$_3$ (001) (STO) substrates by the pulsed laser deposition (PLD) technique. The Nb-doped substrates were used to avoid charging effects in the photoemission measurements. The deposition temperature and oxygen pressure were controlled at 800 °C and 1×10$^{-3}$ mTorr, respectively. The whole PLD growth process was monitored with *in-situ* reflection high-energy electron diffraction (RHEED) to check the growth model and control the film thickness. Finally, the as-prepared films were annealed at 400 °C for 30 min in an atmospheric pressure of oxygen to remove oxygen vacancies, and then cooled down to room temperature. The crystalline phases of thin films were identified by X-ray diffraction (XRD). Atomic force microscope (AFM) topographic measurements were performed on all the measured films samples to observe the surface morphology. Resonant photoemission spectroscopy (RPES) and x-ray absorption spectroscopy (XAS) measurements were performed on BSTO films prepared *in situ* to investigate their electronic structures. The temperature was varied from room temperature (RT) to 12 K. The RPES spectra were measured using a spectrometer equipped with a high-energy-resolution electron energy analyzer Scienta SES-2002. The XAS spectra were measured using the total-electron-yield method. All of experiments were carried out at BL23SU in Spring-8, Japan.

The atomic force microscope (AFM) and reflection high-energy electron diffraction (RHEED) images (not shown here) indicated that the BSTO film was atomically smooth with only single-unit cell scale roughness arising from the step-and-terrace structure of the original substrates, as expected for a layer-by-layer growth. X-ray diffraction (XRD) peaks were observed only from BSTO (00$l$) and the STO (00$h$) substrate as shown in Fig. 1, indicating that the BSTO films are highly oriented in the (001) planes. Figure 2 shows the Ti $L_{2,3}$ edge X-ray absorption (XAS) spectra of the 200 nm-thick BSTO film at room temperature. The first two small peaks are related to a transition which is forbidden in the *LS*-coupling but becomes allowed due to the multipole *p-d* interactions. The third and fourth peaks are ascribed to transitions to the $\underline{2p}_{3/2}3d$ ($t_{2g}$) and $\underline{2p}_{3/2}3d$ ($e_g$) final states of the $L_3$ edge, respectively, where the underlines denote core holes. The fifth and sixth peaks are transitions to the final $\underline{2p}_{1/2}3d$ ($t_{2g}$) and $\underline{2p}_{1/2}3d$ ($e_g$) states of the $L_2$ edge.

RPES was used to isolate the hybridized Ti 3*d*-ligand (oxygen 2*p*) components in the valence band. As the temperature decreases, BSTO relaxes from the high-temperature cubic structure to the lower-symmetry tetragonal structures (the phase transition temperature is about 230 K for bulk $Ba_{0.5}Sr_{0.5}TiO_3$).[8] The displacement of the titanium ions with respect to the center of the oxygen octahedra occurs along the [100] direction, and thus there are two different kinds of Ti-O bond lengths due to the displacement of the Ti atom from the center of the oxygen octahedral. According to theoretical calculations,[10, 12] Ti-O hybridization is essential for ferroelectricity in the perovskite Ti-O oxides compounds, and is intimately connected with the paraelectric to ferroelectric structure changes. Therefore, we expect differences in the shape and strength of RPES spectra in the valence-band region in going through the cubic-to-tetragonal phase transition as observed by Higuchi *et al.*[11]

From the XAS results shown in Fig. 2, we chose $h\nu$ = 466 eV (the Ti 2*p* → 3*d* X-ray absorption transition) as the photon energy to monitor changes in the RPES spectra in the valence-band region of the BSTO films between room temperature and 12 K. Figure 3a shows typical RPES spectra of 200 nm-thick BSTO film taken at $h\nu$ = 466 eV, which have been normalized to the same integrated area after removing the integral background. The results indicate that the valence-band spectra have two main features, i.e., A (4-5.5 eV) and B (6-9 eV), corresponding to the O 2*p* non-bonding and O 2*p*-Ti 3*d* bonding states, respectively, in good agreement with the previous first-principle electronic structure calculations.[13] The splitting of the two-peak structure of the valence band was about 2.6 eV, and is almost the same as that of the valence band observed in a $BaTiO_3$ single crystal.[14] By comparing the spectra of the ferroelectric phase (at 12 K) with the paraelectric phase (at 300 K), there is a remarkable difference in the intensity ratio B/A between the O 2*p* non-bonding region (feature A) and the bonding region (feature B) before and after the phase transition. It is of interest to note that the B/A ratio is larger, that is, the Ti-O hybridization in the cubic paraelectric phase is stronger than that in the ferroelectric phase, which was also observed in the Ti 3*p* → 3*d* RPES experiment by Higuchi and his co-workers for bulk $BaTiO_3$ crystal by changing the measurement temperature.[11]

We have also measured the RPES spectra in the valence-band region of the other three BSTO films of different thicknesses as shown in Fig. 3b-3d, and found that the 2.8 nm-thick BSTO film also exhibited such changes similar to those of the 200 nm-thick BSTO film. However, the spectra of 2.0 nm and 1.2 nm BSTO films changed only weakly at the different measurement temperatures. We define Δ as the ratio of the absolute spectral height difference to the maximum spectral peak height, and plot it as a function of the film thickness. As clearly seen, the Δ value for the 200 nm-thick film (or 2.0 nm-thick film) is very close to that for the 2.8 nm-thick film (or 1.2 nm-thick film), but the Δ value decreases abruptly from 2.8 nm to 2.0 nm. This sharp drop in the Δ value implies that there exists a critical thickness for phase change of the BSTO films in the thickness range of 2.0-2.8 nm. Here, it should be cautioned that the reduction of the temperature dependent change with decreasing BSTO film thickness may partly be due to the decreasing contribution of the BSTO and the increasing contribution of the STO substrate. In order to estimate this extrinsic effect, we consider the effect of thinning of BSTO film thickness on the BSTO signal intensity as:

$$I = 1 - \exp(-d/\lambda) \tag{1}$$

where $d$ is the thickness of the BSTO film, and $\lambda \sim 10$ Å is the photoelectron mean-free path at the photon energy of 466 eV.[15] Figure 4 shows the comparison of experimental intensity change and the calculated intensity change using Eq. (1). The figure clearly shows that the experimental intensity decreased more quickly than the calculated one with decreasing film thickness below 2.8 nm, which implies that the decrease in the experimental results should be an intrinsic effect caused by the phase change with decreasing film thickness, and cannot be explained simply by the effect of the thinning of the BSTO film. It is noted that Junquera and Ghosez[16] have ever predicted that $BaTiO_3$ films will lose their ferroelectric properties below a critical thickness of about 2.4 nm.

However, this critical size effect for the ferroelectric phase may also be related to extrinsic factors such as misfit strain, defects and depolarization field.[17] Our epitaxial BSTO thin film was annealed at 400 °C for 30 min in an atmospheric pressure of oxygen to remove oxygen vacancies. The reciprocal space maps (not presented here) of XRD data indicated that the coherent growth of BSTO film on the STO substrate was observed. The misfit strain existing in these BSTO films grown on the STO substrate can be evaluated on the basis of a phenomenological thermodynamic theory. As Ban and his co-worker's calculations,[18] the misfit strain did not change abruptly for these 2.8 nm and 2.0 nm BSTO films on STO substrate. Therefore, this electronic structure change should be mainly ascribed to the intrinsic size effects, which results in the critical thickness for the phase change of the BSTO thin films.

In summary, the resonant photoemission valence-band spectra of the epitaxial BSTO films have been measured at RT and 12 K. Our results indicate that there exists a phase change for the BSTO films in the thickness range of 2.0-2.8 nm, which should be ascribed to the intrinsic critical size effects. Measurements of BSTO ultrathin films on different substrates (e.g., MgO, LaAlO$_3$, DyScO$_3$) to understand the misfit strain effect on the phase changes using the present method are desirable.

## Acknowledgements

The experiment was performed under the approval of the Japan Synchrotron Radiation Research Institute (JASRI) Proposal Review Committee (Proposal No. 2005A0285-NSa-np-Na). This work was partially supported by JSPS (A19204037) and the NSF of China (Grant No. 50621201 and 10574078). YHL thanks JSPS fellowship for Foreign Researchers and G. Liu for his valuable discussions.

**References:**

1. J. F. Scott, Ferroelectric Memories (Springer-Verlag, New York, 2000).

2. W. L. Zhong, Y. G. Wang, P. L. Zhang, B. D. Qu, Phys. Rev. B 50, 698 (1994).

3. M. Chu, I. Szafraniak, R. Scholz, C. Harnagea, D. Hesse, M. Alexe, U. Gösele, Nature Mater. 3, 87 (2004).

4. A. Roelofs, T. Schneller, K. Szot and R. Waser, Nanotechnology 14, 250 (2003).

5. D. D. Fong, G. B. Stephenson, S. K. Streiffer, J. A. Eastman, O. Auciello, P. H. Fuoss, C. Thompson, Science 304, 1650 (2004).

6. G. Gerra, A. Tagantsev, N. Setter, K. Parlinski. Phys. Rev. Lett. 96, 107603 (2006).

7. Y. S. Kim, D. H. Kim, J. D. Kim, Y. J. Chang, T. W. Noh, J. H. Kong, K. Char, Y. D. Park, S. D. Bu, J.-G. Yoon, J.-S. Chung, Appl. Phys. Lett. 86, 102907 (2005).

8. V. V. Lemanov, E. Smirnova, P. Syrnikov, E. Tarakanov, Phys. Rev. B 54, 3151 (1996).

9. S. W. Robey, L. T. Hudson, V. E. Henrich, C. Eylem, B. Eichhorn, J. Phys. Chem. Solid. 57, 1385 (1996).

10. R. E. Cohen, H. Krakauer, Phys. Rev. B 42, 6416 (1990).

11. T. Higuchi, T. Tsukamoto, K. Oka, T. Yokoya, Y. Tezuka, S. Shin, Jpn. J. Appl. Phys. 38, 5667 (1999).

12. R. E. Cohen, Nature 358, 136 (1992).

13. P. Pertosa, G. Hollinger, F. M. Michel-Calendini, Phys. Rev. B 18, 5177 (1978).

14. B. Cord, R. Courths, Surf. Sci. 152/153, 1141 (1985).

15. S. Tanuma, C. J. Powell, D. R. Penn, Surf. Sci. 192, L849 (1987).

16. J. Junquera. P. Ghosez, Nature 422, 506 (2003).

17. M. Dawber, K. M. Rabe, and J. F. Scott, Rev. Mod. Phys. 77, 1083 (2005).

18. Z. -G. Ban and S. P. Alpay. J. Appl. Phys., 93, 504 (2003).

**Figure captions:**

Fig. 1    XRD patterns of the 200 nm-thick BSTO film grown on the Nb-doped STO (001) substrate.

Fig. 2    Typical XAS spectra of the 200 nm-thick BSTO film at room temperature.

Fig. 3    Resonant photoemission spectra in the valence-band region of BSTO films with various thicknesses grown on Nb-doped STO substrates (a-d).

Fig. 4   Comparison of experimental intensity change and the calculated intensity change assuming only the effect of the thinning of BSTO film. Vertical bars are deviation scale

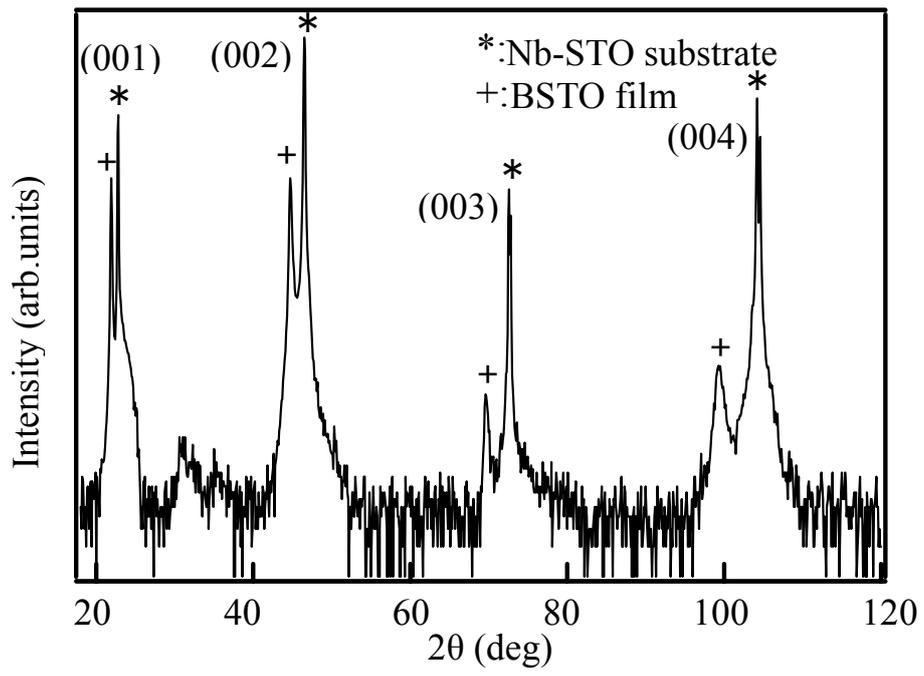

Fig. 1

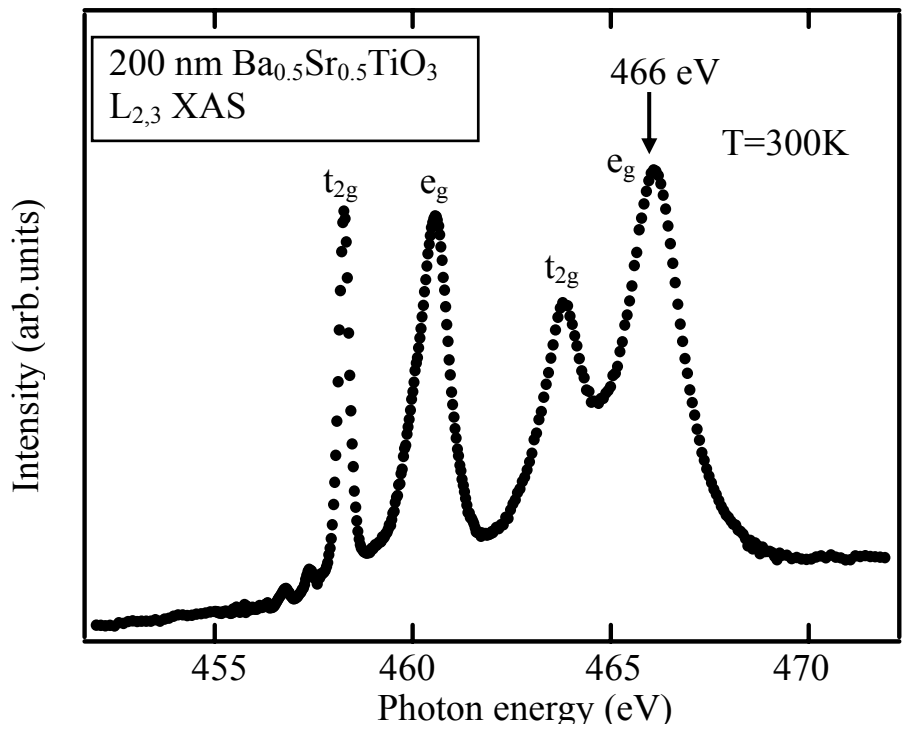

Fig. 2

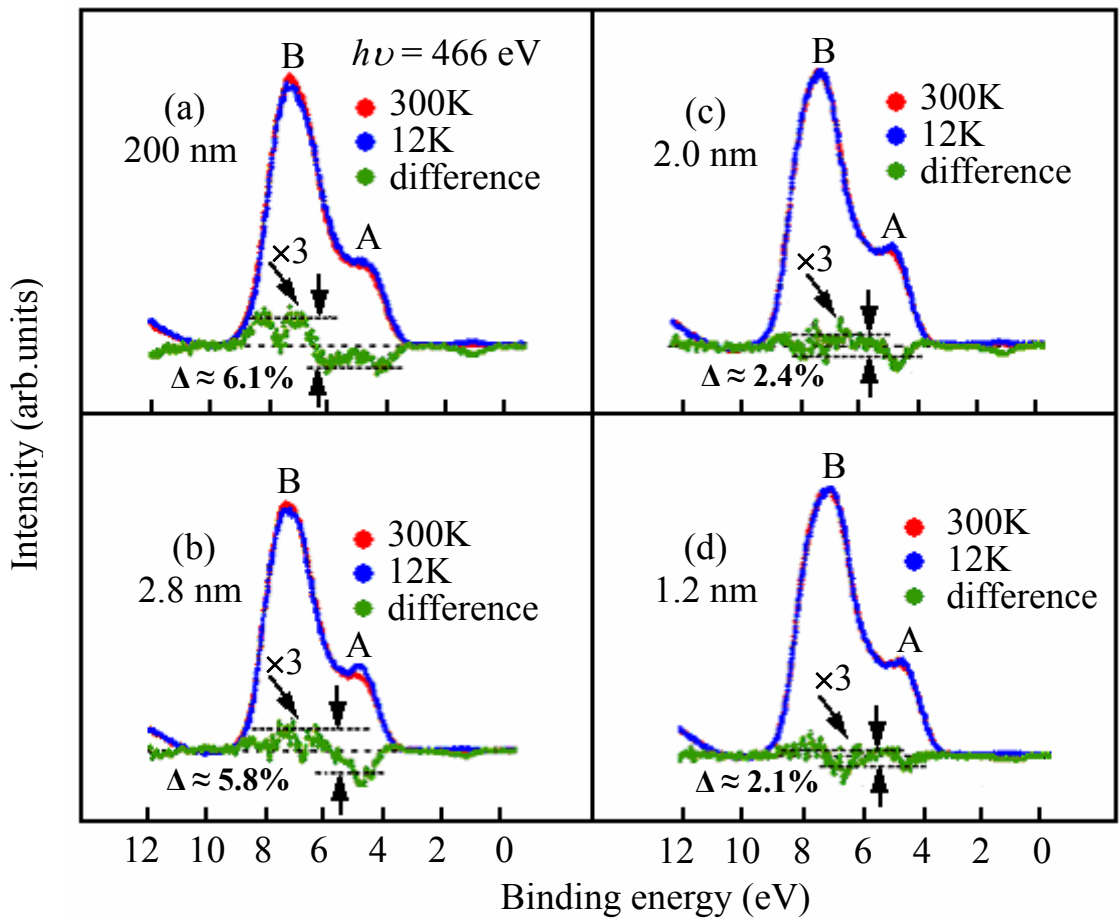

Fig. 3

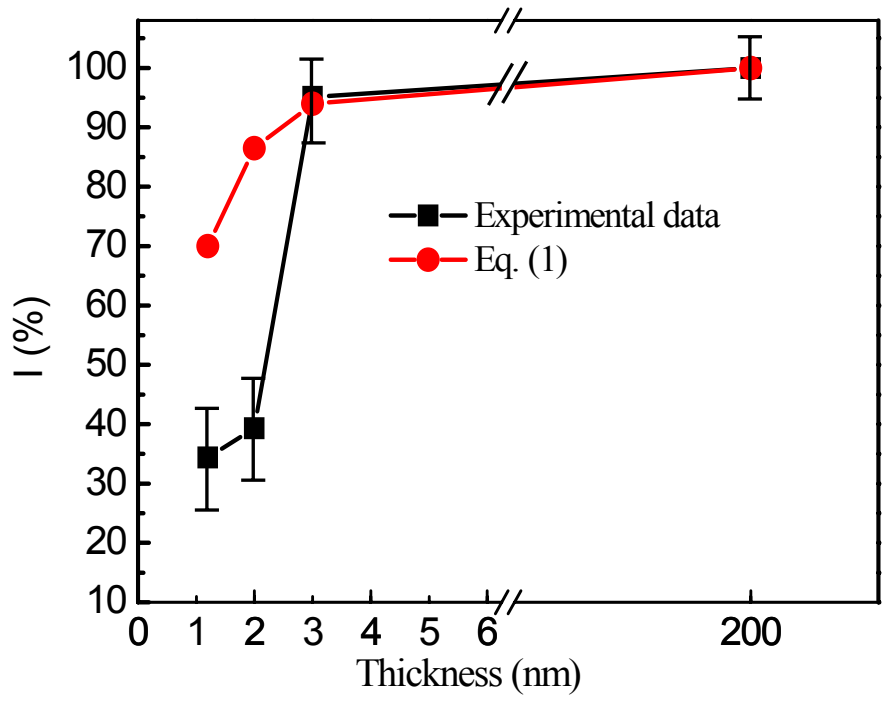

Fig. 4